\documentclass[letterpaper,english,preprint, aps]{revtex4-1}
\usepackage[T1]{fontenc}
\usepackage[latin9]{inputenc}
\usepackage{float}
\usepackage{amsmath}
\usepackage{amssymb}
\usepackage{graphicx}

\makeatletter

\pdfpageheight\paperheight
\pdfpagewidth\paperwidth

\usepackage{braket}
\usepackage{bbm}

\makeatother

\usepackage{babel}
\begin{document}
\preprint{BROWN-HET-1784}
\title{Black Hole Interiors via Spin Models}
\author{David A. Lowe}
\email{lowe@brown.edu}

\author{Mengyang Tong}
\email{mengyang_tong@brown.edu}

\affiliation{Department of Physics, Brown University, Providence, RI, 02912, USA}
\begin{abstract}
To model the interior of a black hole, a study is made of a spin system
with long-range random four-spin couplings that exhibits quantum chaos.
The black hole limit corresponds to a system where the microstates
are approximately degenerate and equally likely, corresponding to
the high temperature limit of the spin system. At the leading level
of approximation, reconstruction of bulk physics implies that local
probes of the black hole should exhibit free propagation and unitary
local evolution. We test the conjecture that a particular mean field
Hamiltonian provides such a local bulk Hamiltonian by numerically
solving the exact Schrodinger equation and comparing the time evolution
to the approximate mean field time values. We find excellent agreement
between the two time evolutions for timescales smaller than the scrambling
time. In earlier work, it was shown bulk evolution along comparable
timeslices is spoiled by the presence of the curvature singularity,
thus the matching found in the present work provides evidence of the
success of this approach to interior holography. The numerical solutions
also provide a useful testing ground for various measures of quantum
chaos and global scrambling. A number of different observables, such
as entanglement entropy, out-of-time-order correlators, and trace
distance are used to study these effects. This leads to a suitable
definition of scrambling time, and evidence is presented showing a
logarithmic variation with the system size.
\end{abstract}
\maketitle

\section{Introduction}

The Anti-de Sitter/Conformal Field Theory correspondence (AdS/CFT)
\citep{Aharony:1999ti} has been tremendously successful in providing
a framework for addressing questions in quantum gravity, that goes
far beyond the successes of perturbative string theory. In particular
it provides a detailed accounting of black hole entropy and important
information about the nonperturbative vacuum structure of string theory/quantum
gravity. The holographic mapping from conformal field theory operators
to bulk spacetime operators is however only well-understood as a perturbative
expansion around asymptotically AdS regions \citep{Hamilton:2005ju,Hamilton:2006az},
and there is much current debate about how (or even whether) the holographic
mapping can be extended deep into the bulk spacetime, where the presence
of apparent horizons and global horizons make the application of the
perturbative holographic mapping problematic.

To make progress on these issues, it is necessary to develop an understanding
of the holographic mapping that is less dependent on the special conformal
symmetry of AdS, and instead can work in much more general backgrounds.
In a series of papers, it has been argued a more general holographic
mapping should take the form of a mean field theory approximation,
where the bulk degrees of freedom are to be extracted after suitable
averaging over the microscopic exact representation \citep{Lowe:2014vfa,Lowe:2015eba,Lowe:2016mhi,Lowe:2017ehz}.
In some sense, this is not a new idea, and similar proposals have
been made in the context of loop quantum gravity, fuzzballs, etc.
However what is new about the current work is that a specific class
of Hamiltonia are proposed to describe black hole interiors, and a
specific form of the mean field approximation is developed that may
then be tested in detail.

In earlier work \citep{Lowe:2017ehz}, this idea was developed for
the simpler case of a general (typically long-range, random) two-spin
interaction. However the simple form of the interaction opens the
door to questions about whether such systems can really exhibit quantum
chaos. Since the systems have finite dimensional Hilbert spaces, by
construction, the Hamiltonian may always be diagonalized, and questions
of chaos boil down to whether the spectrum of energy eigenvalues is
suitably dense, and whether the ``local'' basis of states one might
be interested in have a simple representation in terms of energy eigenstates.
The latter condition is not satisfied if we restrict to interactions
with a range comparable to the system extent. So it remains to ensure
the energy spectrum is not too sparse as to induce time recurrences.
This effect emerged as a feature in some of the toy model calculations
of \citep{Lowe:2017ehz}, and in part motivates the present work.
By considering a four-spin interaction, the system is expected to
exhibit quantum chaos, with recurrences only expected on timescales
parametrically larger than the scrambling time. However, in addition,
the couplings will be chosen to follow a random distribution, ensuring
the spectrum of energy eigenvalues is suitably dense.

The starting point for translating states in such a description to
bulk states is to suppose that at some given time one can pick a basis
corresponding to bulk fields localized on some shell of fixed proper
radius in the vicinity of the horizon of a black hole. For the present
work we will not consider additional charges, nor rotation and presume
we have a simple Schwarzschild black hole. For now, the number of
spacetime dimensions will be left arbitrary. Such a shell can be viewed
as our holographic screen, and for bulk excitations localized on such
a shell, the holographic map will be particularly simple. As time
evolves, the excitations will move forward in time, typically to smaller
radii. In the limit of a large black hole, we expect these test probes
to follow timelike geodesics ending on the singularity.

We assume a good approximation to this choice of basis can be made,
which amounts to conjugating some initial Hamiltonian by a unitary
transformation. The main physical assumptions we make are that this
Hamiltonian exhibits fast scrambling, in a sense to be defined below,
and that the Bekenstein-Hawking entropy of the black hole is chosen
to match the log of the Hilbert space dimension
\begin{equation}
S_{BH}=N\log2\,.\label{entropyrel}
\end{equation}
For the purposes of the numerics below we further assume a random
four-spin Hamiltonian is sufficiently general to capture the relevant
properties, with the black hole is represented by some randomly chosen
vector in the Hilbert space of a large number $N$ of spins,

To represent a bulk probe, a smaller system of spins is tensored to
this Hilbert space and a pure state is constructed in this Hilbert
subspace. The full state is then a product of two pure states. Under
the exact time evolution, these states become entangled, and the reduced
density matrix in the probe Hilbert subspace becomes mixed. However
this exact evolution is at odds with what is expected from time evolution
of bulk fields with respect to a local Lagrangian. The primary goal
of the present work is to test the idea that this mixing can be neglected
for times less than the scrambling time, and explore a variety of
measures that are diagnostics of this mixing. This in turn will lead
to a definition of the global thermalization (or scrambling) time
as we detail below, and we will see that even in this high temperature
limit, the spin model exhibits a version of fast scrambling, where
the scrambling time is logarithmic in the system size.

\section{Holographic Model}

The basic form of the holographic model to be studied in the present
paper is a $N$ spin-$1/2$ system with a 4-spin non-local coupling
\begin{equation}
H=\sum_{1\leq i<j<k<l\leq N}J_{ijkl}\overrightarrow{s}_{i}\overrightarrow{s}_{j}\overrightarrow{s}_{k}\overrightarrow{s}_{l}\label{fourspinham}
\end{equation}
where $J_{ijkl}$ are random couplings proportional to $J$ times
a unit normal distribution, completely symmetric in the indices $i,j,k,l$
and pointing in a random direction in spin space. The $\overrightarrow{s}_{i}$
are spin-$1/2$ operators acting on site $i$ where $i=1,\cdots,N$.
The limit of interest is to take $N$ large while keeping 
\begin{equation}
\mathrm{var}(H)=N^{0}\label{variance}
\end{equation}
where the variance is taken over the full Hilbert space. This corresponds
to taking
\[
J^{-2}=\left(\begin{array}{c}
N\\
4
\end{array}\right)\,.
\]
This condition will be discussed further below. We note this large
$N$ limit is different than the SYK model, where instead one scales
for fast scrambling at low temperatures, where an additional conformal
symmetry emerges, and the holographic dual includes an entire AdS
asymptotic region. High temperature scrambling in this (and a more
general class of such models) has been studied in \citep{Bentsen:2018uph}.

In practice, we will perform numerical computations at finite $N$,
imposing the condition \eqref{variance} to normalize the couplings
$J_{ijkl}$. One of our goals will be to show that such a system fast
scrambles in the limit of large temperature, in the sense that the
scrambling time to be defined below behaves as 
\begin{equation}
t_{scr}\sim N^{0}\log N\label{tscram}
\end{equation}
We note in this theory nothing yet depends on the dimensionality of
spacetime. The theory is intended to reproduce the correct chaotic
physics of the horizon for timescales of order the scrambling time.
In future work, local dimension dependent modifications of the theory
will be studied which can allow the holographic mapping to the bulk
to be further completed, with a view to including the correct local
interactions, beyond geodesic propagation.

\subsection*{Matching with the bulk}

The theory in question may be viewed as a model for the stretched
horizon theory, in the sense of \citep{Thorne:1986iy}. For simplicity
we consider a Schwarzschild black hole in general spacetime dimension
$D$
\[
ds^{2}=\left(1-\frac{2M}{r^{D-3}}\right)dt^{2}-\left(1-\frac{2M}{r^{D-3}}\right)^{-1}dr^{2}+r^{2}d\Omega_{D-2}^{2}
\]
where we work in units where $G=1=c=\hbar$. Since we will match the
Bekenstein-Hawking entropy with $N$ via \eqref{entropyrel}, it will
be helpful to tabulate the $N$ dependence of the thermodynamic observables
of the black hole
\[
M\sim N^{\frac{D-3}{D-2}},\,T_{BH}\sim N^{-\frac{1}{D-2}},\,r_{H}\sim N^{\frac{1}{D-2}},\,\left(\delta M\right)^{2}=T_{BH}^{2}\frac{\partial M}{\partial T_{BH}}\sim-N^{\frac{D-4}{D-2}}
\]
With respect to Schwarzschild time, which measures proper time near
$r=\infty$, the energy is simply $M$. The set of states associated
with the microscopic Hamiltonian $H$ will split this energy into
a band of states and we choose to normalize the width of this band
according to the following relation
\begin{equation}
E=M(N)+N^{-\frac{1}{D-2}}H\label{microenergy}
\end{equation}
where $M(N)$ represents the classical black hole mass, which may
be treated as an $N$ dependent constant shift in the Hamiltonian.
This scaling corresponds to choosing to define the position of the
stretched horizon such that the redshift with respect to infinity
converts a Planck energy down to an energy equal to the Hawking temperature.

With this scaling, we recover the expected expression \citep{Sekino:2008he}
for the scrambling time in Schwarzschild coordinates from the relation
\eqref{tscram}
\[
t_{scr,S}\sim\frac{1}{T_{BH}}\log S_{BH}
\]
One can also look at the contribution of the microscopic Hamiltonian
to the width of the energy spectrum
\begin{equation}
(\delta E)^{2}\sim-N^{\frac{D-4}{D-2}}+\mathcal{O}(N^{-\frac{2}{D-2}})\label{energyfluc}
\end{equation}
where the first term is the semiclassical result, reflecting the negative
specific heat of the black hole, and the second term is due to the
width of the microscopic spectrum of spin states. We see this extra
width matches the energy scale associated with a single Hawking particle
of energy $T_{BH}$, which is a physically reasonable result. It is
also compatible with treating the holographic model in the high temperature
limit, since the physical temperature $T_{BH}$ induces a much larger
scale for fluctuations via the semiclassical term (the first term
in \eqref{energyfluc}) versus the term arising from the microscopic
Hamiltonian. Our goal is to use the Hamiltonian $H$ to model the
dynamics of the black hole on timescales below the Page time ($t_{Page}\sim N^{\frac{D-1}{D-2}}\gg t_{S})$,
when we may approximate the black hole as having an a constant mass
(and hence $N$).

To derive black hole thermodynamics from the microscopic energy \eqref{microenergy}
one can use the relation $S=\log\Omega$ where $\Omega$ is the dimension
of the Hilbert space of the spin model. This statement implicitly
assumes one is coarse graining over energies of order $T_{BH}$, allowing
one to simply count all the states in the spin model. With such a
coarse graining, one may also drop the second term in \eqref{microenergy}.
Applying the usual rules of thermodynamics to $S(E)$ then reproduces
all the expected relations of black hole thermodynamics.

In general, the holographic mapping to the dual gravity variables
is expected to be highly non-local. However in the present situation,
we can take advantage of the infinite range interactions to simplify
this map. At any given time, we can try to perform a general unitary
transformation on our $2^{N}$dimensional Hilbert space to organize
the space into $N$ sites with a spin $1/2$ degree of freedom at
each site, where each site is to be thought of as some point on a
sphere of some fixed radius inside or near the horizon. As time progresses,
this map will become much more complicated, but we will be chiefly
interested in the early time behavior, prior to the scrambling time
$t_{scr}$, where this quasi-local bulk interpretation of the Hilbert
space is a good approximation. The immediate goal then is to build
a candidate for the bulk Hamiltonian, which is local, and which provides
an alternative evolution to the exact time evolution, which is essentially
indistinguishable in this range of times.

Now there is no guarantee this procedure will work, chiefly because
we have little insight into the detailed form of the correct interactions,
and whether this quasi-localization basis can actually be constructed
for the Hamiltonian that descends from some complete theory of quantum
gravity. But as we will see in the present work, the results are largely
insensitive to the details of the interaction chosen, so a goal of
the present work is to present this picture in as general a context
as possible so it may one day be applied to some correct Hamiltonian
of the black hole.

\begin{figure}[H]
\includegraphics[scale=0.7]{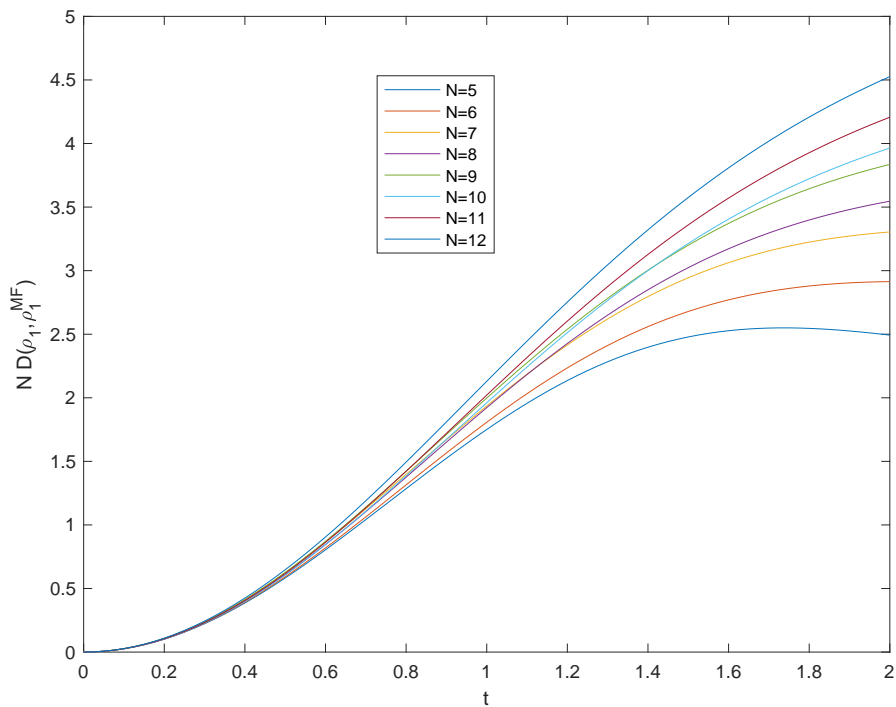}

\caption{\label{fig:ntrdist}$N\ D\left(\rho_{1}(t),\rho_{1}^{MF}(t)\right)$
for various $N$. For each $N$ , trace distance divergence is averaged
over random Page states and over the ensemble of $H$. In the second
panel, the early time region is shown and compared to $\frac{1}{2}\left(1-\sqrt{2P-1}\right)$
with dashed lines, where the bound \eqref{eq:tbound} is close to
being saturated.}
\end{figure}

\subsection*{Observables}

The Hilbert space is the tensor product of individual spin sites $\mathcal{H}=\otimes_{i=1}^{N}\mathcal{H}_{i}$.
Under the assumption the system exhibits quantum chaos for typical
states, then such a typical state will scramble in a timescale of
order $t_{scr}$ and we can use such a state to describe a black hole.
In practice, we will simply choose a random unitary vector in the
Hilbert space $\mathcal{H}$ to generate candidate black hole states.

To represent an observer (or test particle) entering the black hole,
we enlarge the Hilbert space (for example taking $N\to N+1$ and begin
in a product state

\begin{equation}
|\psi(0)\rangle=|\psi_{1}\rangle\otimes|\psi_{0}^{bh}\rangle\label{eq:prodstate}
\end{equation}
where $|\psi_{1}\rangle$ is the spin state representing the observer,
and is chosen to be $|\uparrow\rangle$ and $|\psi_{0}^{bh}\rangle$
describes the state of the black hole. The Hamiltonian $H$ generates
the exact time evolution of the system. However in a scrambling time
the observer's spin becomes entangled with the black hole. Certainly
with respect to the original local basis, the observer's reduced density
matrix becomes highly mixed, and they do not experience the expected
laws of quantum mechanics following from time evolution along a bulk
geodesic. On the other hand, the efficient averaging of the maximally
non-local interaction suggests mean field can be appropriate. As we
now see, this leads to evolution that preserves the pure state tensor
structure \eqref{eq:prodstate}, as we would expect for a particle
moving along a bulk geodesic.

\section{Mean Field Versus Exact Evolution}

\begin{figure}
\includegraphics[scale=0.45]{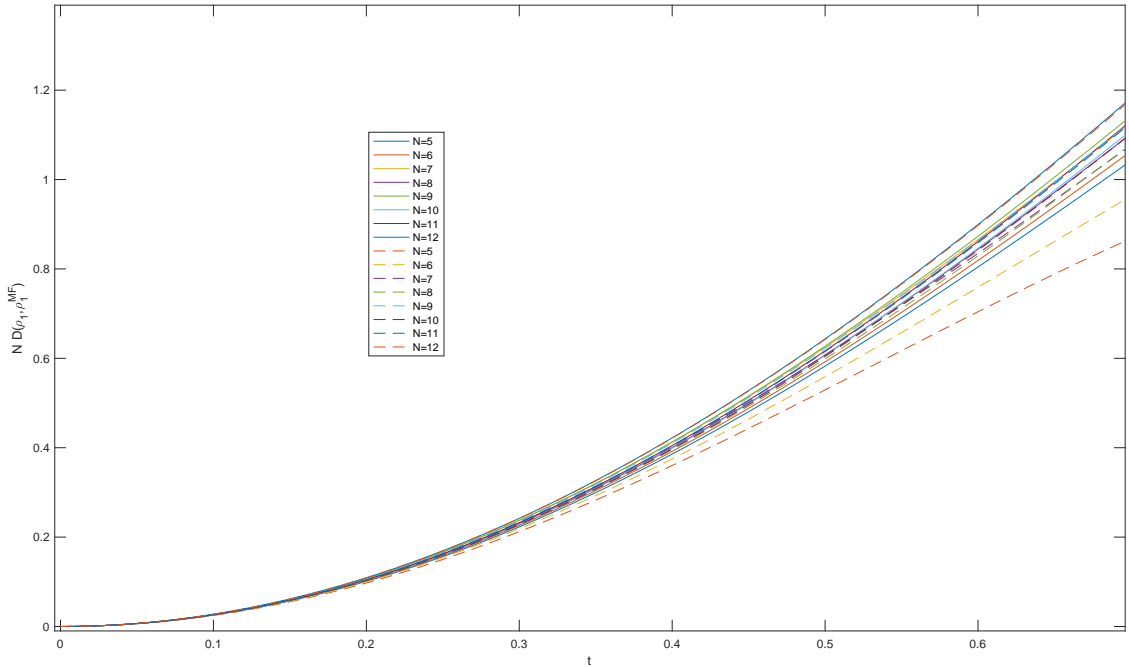}

\caption{The trace distance (solid lines) is compared to the purity bound $\frac{1}{2}\left(1-\sqrt{2P-1}\right)$
(dashed lines). The bound \eqref{eq:tbound} is close to being saturated.}

\end{figure}
The mean field Hamiltonian is defined as
\begin{equation}
H^{MF}(t)=\sum_{i=1}^{N}H_{i}^{MF}(t)=\sum_{i=1}^{N}Tr_{i^{c}}(H\rho^{MF}(t)),
\end{equation}
where $\rho^{MF}(t)$ is the unitary time evolution of the same initial
state evolved by mean field Hamiltonian, and satisfies the von Neumann
equation
\begin{equation}
\dfrac{d\rho^{MF}(t)}{dt}=-i\left[H^{MF}(t),\rho^{MF}(t)\right].\label{eq:meantime}
\end{equation}
Note that each $H_{i}^{MF}$ is a local operator in the Hilbert subspace
associated with spin $i$. The trace is over the complement $i^{c}$
to the Hilbert subspace associated with spin $i$. The mean field
time evolution is therefore guaranteed to preserve the product form
\eqref{eq:prodstate}. However this operator depends on the state
of the other spins, so the time evolution \eqref{eq:meantime} is
inherently nonlinear.

The mean field state will deviate from the exact time evolution. To
measure this deviation, one can consider the trace distance between
reduced density matrices
\[
D(\rho_{1}(t),\rho_{1}^{MF}(t))\equiv\frac{1}{2}\left\Vert \rho_{1}(t),\rho_{1}^{MF}(t)\right\Vert _{1}=\frac{1}{2}\mathrm{Tr}\,\sqrt{\left(\rho_{1}(t)-\rho_{1}^{MF}(t)\right)^{2}}
\]
which offers a metric to measure distinguishability between two quantum
states. In the case $\rho_{1}^{MF}(t)$ is a pure state, as is the
case here, the trace distance is bounded from below by $\dfrac{1}{2}(1-\sqrt{2P-1})$
where $P$ is the purity of $\rho_{1}(t)$, as shown in appendix \ref{sec:Bounding-the-Trace}.

Trace distance may also be bounded from above by a Lieb-Robinson bound
\citep{Lashkari:2011yi,Lowe:2017ehz}. Combining the purity bound
with Lieb-Robinson bound we obtain

\begin{equation}
\dfrac{8}{3N}t^{2}<D\left(\rho_{1}(t),\rho_{1}^{MF}(t)\right)<\dfrac{c'}{N}e^{ct}.\label{eq:tracebounds}
\end{equation}
Here $c$ and $c'$ are constants independent of $N$. This is enough
to ensure that the decoherence is an $1/N$ effect. If the exponential
behavior is saturated from early times to times well before the purity
levels off at $1/2$, the scrambling timescale \eqref{tscram} will
emerge, as the timescale over which the trace distance increases to
some fixed fraction (say $10\%$ for example). The trace distance
between these density matrices measures what is usually termed decoherence
of the pure probe state at site 1. However because the interactions
are maximally non-local, in this particular model, we expect this
to also be a good measure of the global thermalization properties
of the system, and hence we will use this method to define our notion
of scrambling time. As we will see later, it corresponds well to other
definitions considered in the literature.

In figure \ref{fig:ntrdist} we plot $N\:D\left(\rho_{1}(t),\rho_{1}^{MF}(t)\right)$
which clearly indicates the universal behavior for early times. Eventually
the trace distance saturates, as the purity of $\rho_{1}(t)$ approaches
its minimum of $1/2$. The time at which this saturation occurs increases
with $N$ as expected from \eqref{eq:tracebounds}.

\begin{figure}

\includegraphics[scale=0.7]{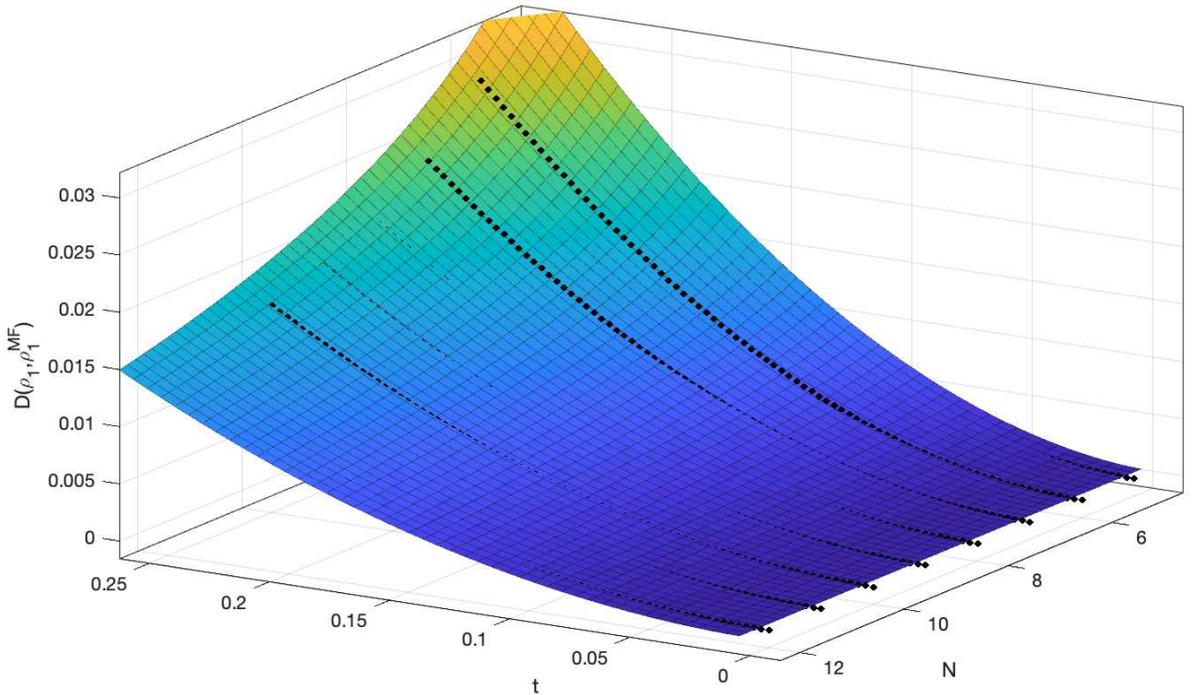}\caption{\label{fig:Fit-of-the}Fit of the averaged trace distance, as a function
of time $t$ and $N$. The surface shows a fit to $at^{2}/N$ where
$a\approx2.6$ in good agreement with the $8/3$ prediction.}

\end{figure}

Following \citep{Yao:2016ayk} one my try a fit of $D\left(\rho_{1}(t),\rho_{1}^{MF}(t)\right)$
to the phenomenologically motivated exponential form
\begin{equation}
D\left(\rho_{1}(t),\rho_{1}^{MF}(t)\right)=a\left(e^{bt}-1\right)^{\Delta}N^{-\gamma}\label{eq:tracefit}
\end{equation}
which yields the values $a=5.5,\,b=0.5,\Delta=1.8$ and $\gamma=.92$.
The value for $\gamma$ is consistent with the analytic bounds \eqref{eq:tracebounds}.
The values for the other parameters are not well determined when looking
at the early time limit. Instead a better fit is obtained simply by
the quadratic form
\[
D\left(\rho_{1}(t),\rho_{1}^{MF}(t)\right)=\frac{a}{N}t^{2}
\]
which yields a better least squares fit with fewer parameters, with
$a=2.6$, agreeing well with the $8/3$ prediction of appendix \eqref{sec:Bounding-the-Trace}.
With this purely quadratic form we do not see evidence of fast scrambling
until we exit this early time limit. In fact, the quadratic approximation
holds very well up until just before the point of inflection in the
curves. This point of inflection then provides one means of defining
the scrambling time. As we see later this matches well with some alternative
measures we study below, which indicate the model does indeed fast
scramble with in a timescale of order $\log N$. In the meantime,
we see the results establish the validity of the mean field approximation
in the window of time prior to the scrambling time.

Rather than study the trace distance between mean field and exact
evolution for a subsystem (in this case a single qubit) we can instead
examine the trace distance between the mean field evolution and the
exact evolution of the full global state over the $N$ qubits. In
line with the expectations of \citep{Hayden:2007cs} we expect a rapid
deviation to emerge. We find a linear increase in the trace distance,
saturating at late times, with behavior largely independent of $N$.

\begin{figure}

\includegraphics[scale=0.7]{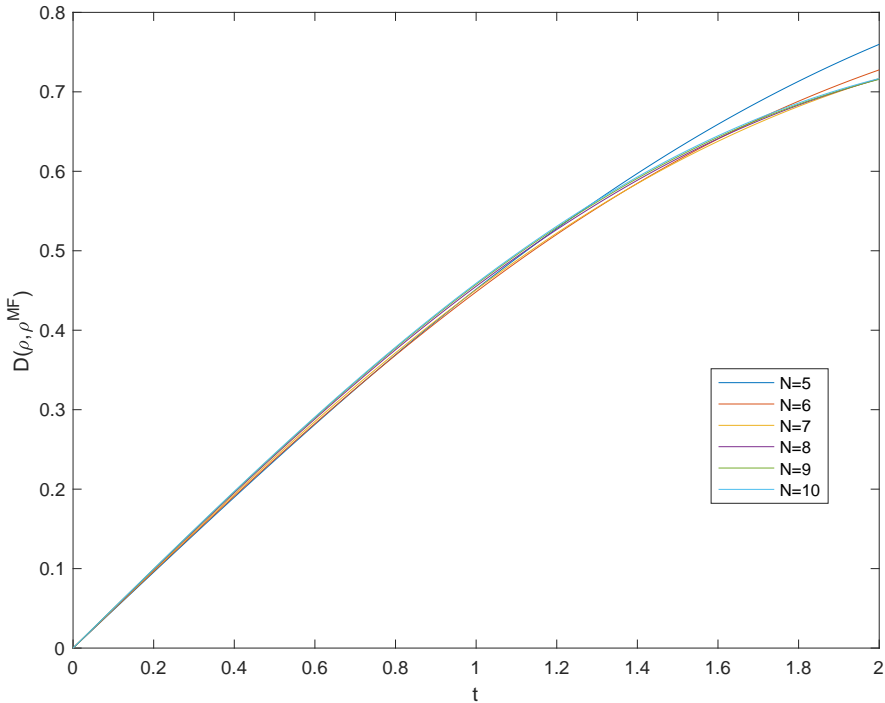}\caption{Trace distance between mean field and exact evolution for the global
state.}

\end{figure}

\begin{figure}
\includegraphics[scale=0.7]{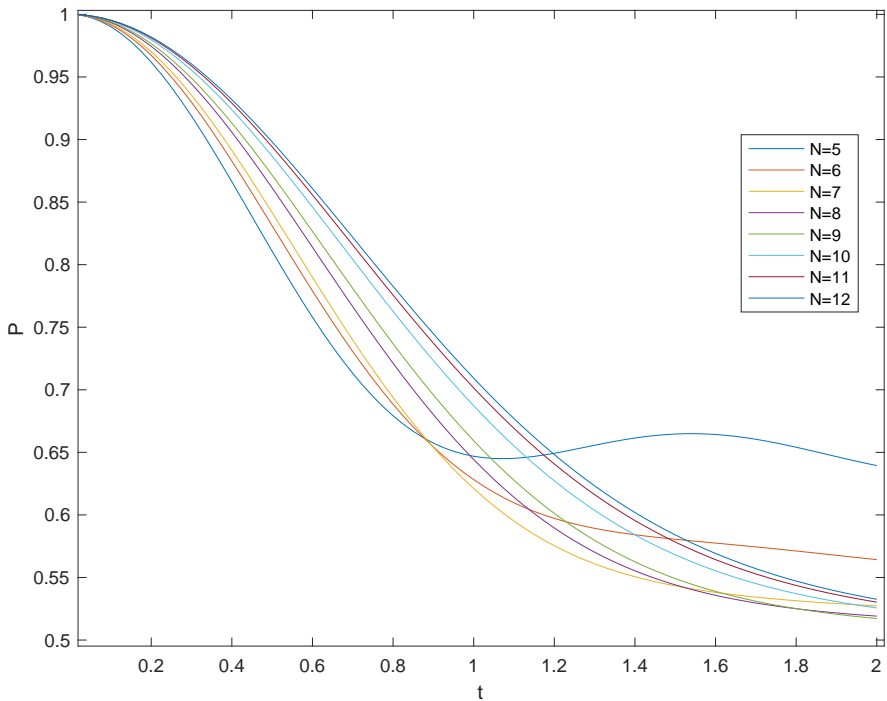}

\caption{\label{fig:Purity}Purity of $\rho_{1}(t)$ for various values of
$N$. For $N>6$ these approach $1/2$ monotonically, as expected
for a system exhibiting quantum chaos.}

\end{figure}

The behavior of the purity of $\rho_{1}(t)$, $P(t)=\mathrm{Tr}\rho_{1}^{2}$
is shown in figure \ref{fig:Purity}. The eigenvalues of the Hamiltonians
are sufficiently dense, due to the choice of random couplings, that
no recurrence is observed over the time range explored. This is an
improvement over \citep{Lowe:2017ehz}, though the early time results
there remain valid despite the simplicity of those models.

Finally we show a fit of the purity as a function of $t$ and $N$,
verifying the form valid for early times found in appendix \ref{sec:Bounding-the-Trace}.
Fitting the form 
\[
P(t)=1-aN^{-\delta}t^{2}
\]
for early times yields $\delta=-0.8$ and $a=-2.8$. Errors with the
expected form arise from the relatively small values of $N$ considered.
Rounding errors also play a role for larger values of $N$.

\begin{figure}
\includegraphics[scale=0.7]{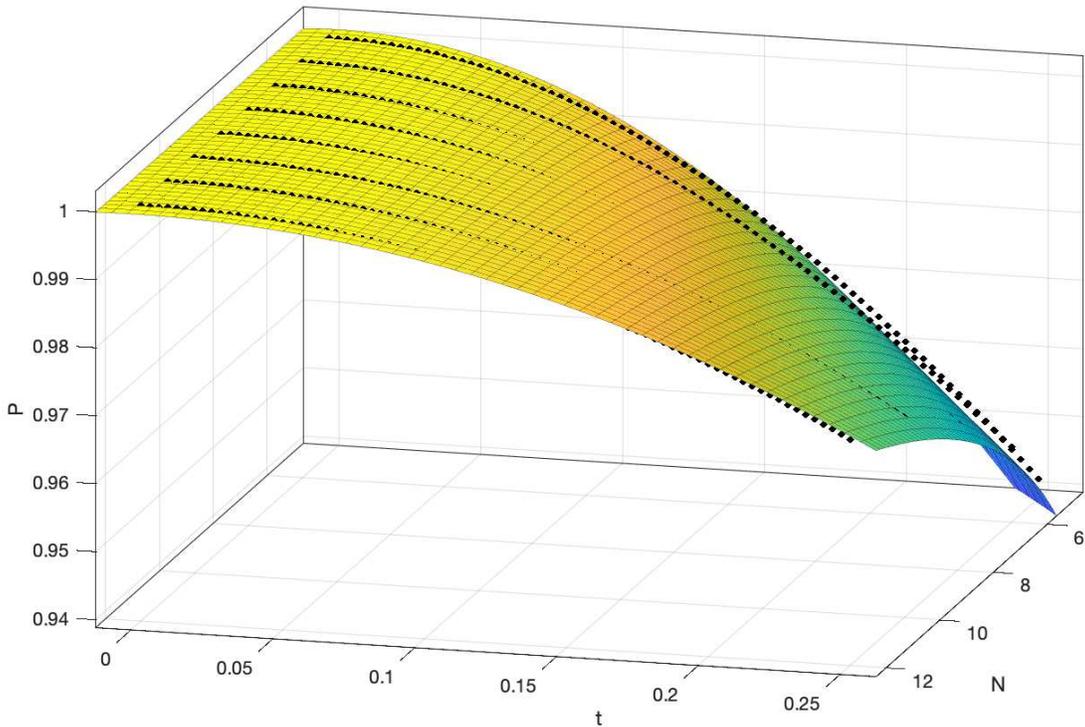}\caption{Purity on the probe site as a function of time $t$ and $N$. Fit
to $1-at^{2}/N$ with $a\approx5.3$ close to the prediction of $16/3$
of appendix \ref{sec:Bounding-the-Trace}.}

\end{figure}
In summary, we have studied numerically and analytically the trace
distance between the mean field and exact evolution for states corresponding
to local probes of black holes in this model. We see the trace distance
remains small for a timescales shorter than the scrambling time \eqref{tscram}.
We have also shown the trace distance is bounded below by the purity,
which depends only on decoherence of a local spin with respect to
the exact time evolution. Moreover this bound is apparently saturated
at early times. These results are consistent with the holographic
interpretation of the mean field as a bulk worldline Hamiltonian advocated
in \citep{Lowe:2017ehz}. At this level of approximation the mean
field approximation is essentially free evolution (more generally
an arbitrary local Hamiltonian can be chosen without changing the
validity of mean field). In future work, we consider adding nearest
neighbor interactions to the spin model to reproduce local field theory
interactions in the bulk. For now we turn to a study of the extent
to which one sees evidence for fast scrambling in this class of models.

\section{Evidence For Fast Scrambling}

The observables studied above are arguably simply studying the thermalization
of the subsystem as interactions place it in contact with the rest
of the system, which acts as the environment, decohering the subsystem.
For a general Hamiltonian, those observables would not be indicative
of global scrambling or quantum chaos. However for the particular
class of Hamiltonians studied here, which are maximally non-local,
we find the results match well with other diagnostics of scrambling
studied in the literature. We now turn to the study of these observables.

\subsection*{OTOC}

The Out-Of-Time-Order-Correlator (OTOC) is one of the first studied
diagnostics of quantum chaos \citep{larkin1969quasiclassical}, where
it was noticed that chaotic dynamics can lead to exponential variation
in such quantities. In particular, the expectation value of the square
of the commutator of a pair of Hermitian operators $V$ and $W$,
$C_{2}(t)=\langle[W(t),V(0)]^{\dagger}[W(t),V(0)]\rangle$ is expected
to grow exponentially in time: $C_{2}(t)\sim e^{\lambda_{L}t}$. Here
$\lambda_{L}$ is to be identified as an analog of a Lyapunov exponent.
Here we study this quantity where $V=s_{z,1}$ and $W=s_{z,2}$.

\begin{figure}[H]
\includegraphics[scale=0.7]{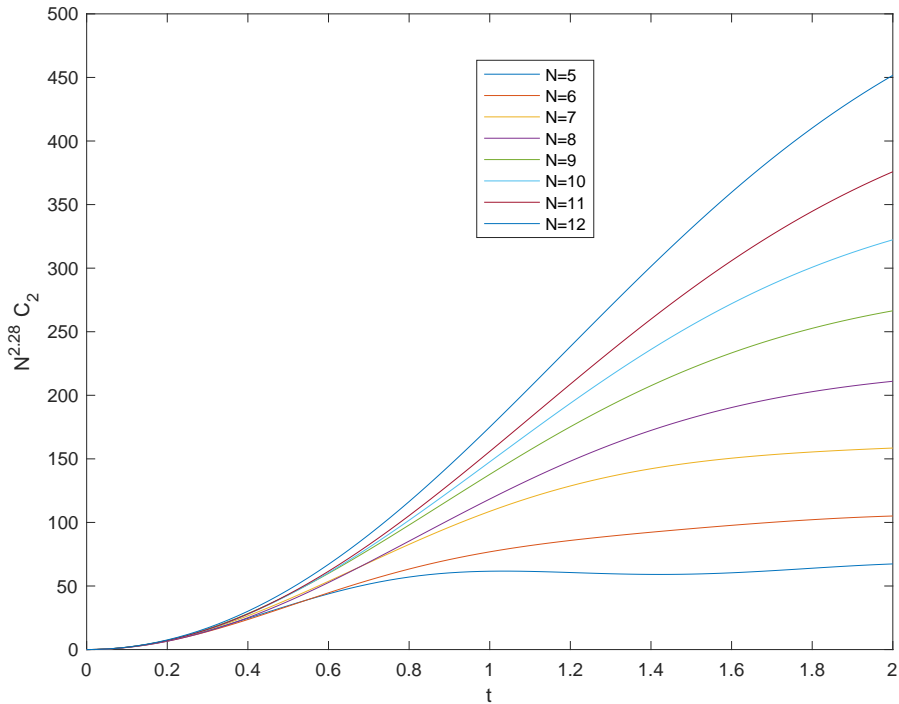}

\caption{Commutator $C_{2}(t)$ as a function of time for various values of
$N$. The result has been rescaled by $N^{2.28}$ to illustrate the
universal early time behavior, prior to saturation/scrambling in the
late time regime.}
\end{figure}
For early time evolution, each line can be fitted by $at^{2}/N^{\delta}$,
where $a$ and $\delta$ are fitting coefficients.

\begin{figure}
\includegraphics[scale=0.7]{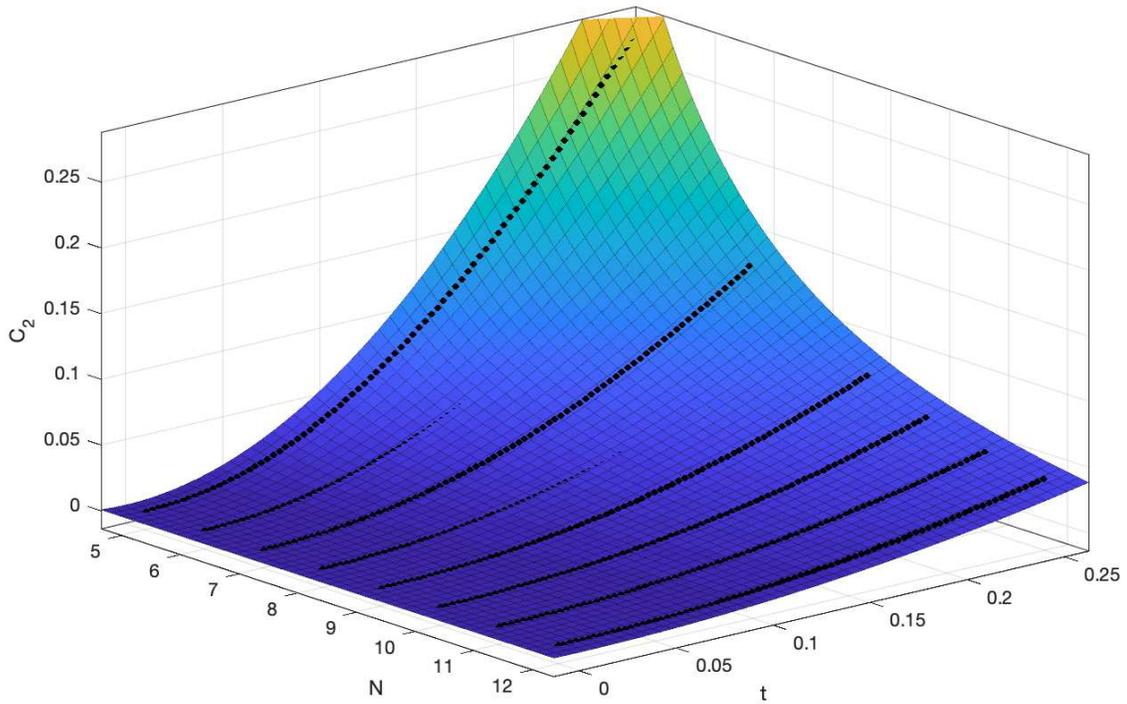}\caption{Here the OTOC is fit to the form $at^{2}/N^{\delta}$ with $a\approx175\pm4,\,\delta\approx2.28\pm0.02$,
showing the early time growth of the commutator.}
\end{figure}

The growth in $C_{2}(t)$ provides the first hint of scrambling. For
the numerically accessible values of $N$ the expected exponential
growth seems to saturate well before there is a clear separation from
the perturbative early time regime (which is only sensitive to the
$t^{2}$ term in an expansion around $t=0$). The behavior of $C_{2}$
is qualitatively very similar to the behavior of the trace distance
(exact vs. mean field) considered in the previous section. To do better
in measuring the scrambling timescale, our strategy will be to perform
a measurement of the cross-over timescale between these different
regimes, and we will study observables where this cross-over can be
measured with greater precision.

\subsection*{Entanglement Entropy}

\begin{figure}[H]
\includegraphics[scale=0.7]{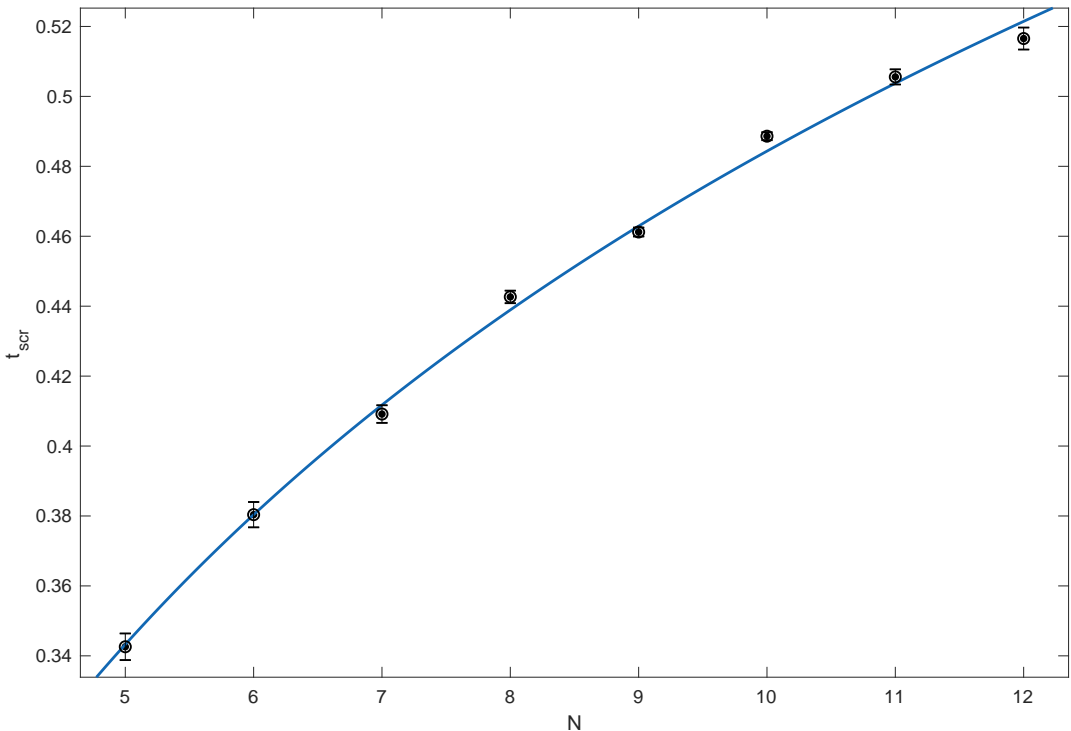}\caption{\label{fig:Scrambling-time-extracted}Scrambling time extracted from
entropy as a function of $N$.}
\end{figure}

Since the Hamiltonian is maximally non-local in the spin basis, we
expect studying the entanglement entropy of pairs of spins provides
a useful measure of the global entanglement of the system, and hence
the extent to which scrambling has taken place. Recalling that we
start the system in a product pure state \eqref{eq:prodstate} we
can compute the entanglement entropy of the spin at site 1, via

\[
S_{ent}(t)=-Tr\left(\rho_{1}\ln\rho_{1}\right).
\]
Below we carry out numerical simulations of the system to obtain entropy
growth as a function of holographic time as shown in figure \ref{fig:Scrambling-time-extracted}.

\begin{figure}[H]
\includegraphics[scale=0.7]{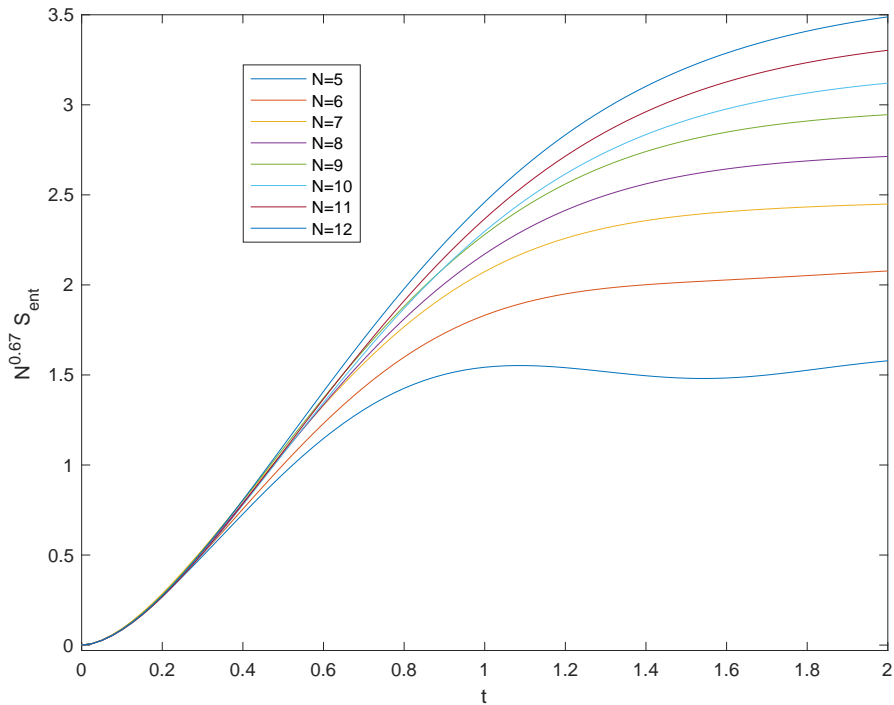}\caption{\label{fig:Entanglement-entropy-as}Entanglement entropy as a function
of time for various $N$. For each $N$ entropy growth is averaged
over random Page states and over the ensemble of $H$. A rescaling
by $N^{.67}$ illustrates the universality of the early time regime.Early
time regime, fit well by $at^{\gamma}/N^{\delta}$ with $a\approx3.7\pm0.1,\gamma\approx1.62\pm0.01,\delta\approx0.67\pm0.01$}
\end{figure}

Entanglement entropy growth in a strongly coupled gapless system with
a gravity dual has been studied intensely in \citep{Casini:2015zua,Liu:2013iza,Liu:2013qca}.
It was proposed that the growth in entanglement entropy can be visualized
as the spreading of an ``entanglement tsunami''. The region covered
by the wave-front is entangled with the rest of the system, and the
system achieves saturation when fully covered by the tsunami. For
a wide class of black hole systems, entropy growth exhibit three stages
of development: 1. pre-local-equilibration quadratic growth; 2. post-local
equilibration linear growth; 3. saturation, exactly matched by our
numeric simulation. It was discussed in \citep{Liu:2013iza} that
the linear growth regime characterizes the late-time memory loss:
the wave front forms a uniform circle and propagates in the same way
regardless of what the initial configuration is. For us, we are interested
in this timescale to reach the linear regime. From the data in fig.
\ref{fig:Entanglement-entropy-as} we can study the position where
the point of inflection of the curves.

The results are shown in figure \ref{fig:Scrambling-time-extracted}.
The timescale shows an obvious $\log N$ dependence $t_{scr}=0.21\log N$.
We note this one parameter fit produces a better fit than a three
parameter power law fit $aN^{\delta}+c$, which offers the strongest
numerical evidence of fast scrambling we have found. The entanglement
entropy does not depend on a choice of operators (as in the OTOC)
nor on the mean field approximation (as in the previous section),
so for us is the most numerically useful quantity to study in the
approach to global scrambling.

\section{Summary}

We have explored a four-spin interacting system that exhibits fast
scrambling feature in the high temperature limit which is conjectured
to be holographically dual to a black hole spacetime in the vicinity
of the horizon. This is in contrast with SYK model, where chaotic
behaviors emerge when the couplings $\beta J$ are taken to be large,
corresponding to a low temperature limit, and holographically to a
complete asymptotically anti-de Sitter spacetime. We extend early
work of the mean field construction to this new holographic Hamiltonian.
The trace distance between the exact and mean field Hamiltonian remain
small for at least a scrambling time which indicates the local mean
field viewpoint, which may be reinterpreted in terms of a bulk description
can be valid for timescales smaller than the scrambling time. This
supports the conjecture that decoherence of the in-falling state is
a dual to the disruptive bulk effects near the space-time singularity.
\begin{acknowledgments}
This work was supported by Brown University through the use of the
facilities of its Center for Computation and Visualization. D.L. is
supported in part by DOE grant de-sc0010010. D.L. acknowledges support
from the Simons Center for Geometry and Physics, Stony Brook University
during the completion of this work. D.L. thanks L. Thorlacius for
discussions.
\end{acknowledgments}

\appendix

\section{Bounding the Trace Distance\label{sec:Bounding-the-Trace}}

For a density matrix $\rho$, the purity is defined by
\begin{equation}
P(t)=\mathrm{Tr}\rho^{2}\,.\label{eq:purity}
\end{equation}
The purity may be computed in an early time expansion following \citep{Kim1996}
\[
P(t)=1-2t^{2}\left(\left\langle H\right\rangle _{\psi}^{2}+\left\langle H^{2}\right\rangle _{\psi}-\left\langle \left\langle H\right\rangle _{\psi_{1}}^{2}\right\rangle _{\psi_{2}}-\left\langle \left\langle H\right\rangle _{\psi_{2}}^{2}\right\rangle _{\psi_{1}}\right)+\mathcal{O}(t^{4})
\]
where we have a pure state $\left|\psi\right\rangle =\left|\psi_{1}\right\rangle \times\left|\psi_{2}\right\rangle $
and use the notation for partial matrix elements $\left\langle H\right\rangle _{\psi_{k}}=\left\langle \psi_{k}\left|H\right|\psi_{k}\right\rangle $.
If we approximate this formula using an average over Page states,
subject to the normalization \eqref{variance} we obtain

\[
P(t)\approx1-\frac{16t^{2}}{3N}+\mathcal{O}(t^{4})\,.
\]
We may use the purity to bound the trace distance 
\[
D(\rho,\rho')=\frac{1}{2}\left\Vert \rho-\rho'\right\Vert _{1}
\]
which is our most refined observable used to define global thermalization.
To do this in the examples studied here, we may use the geometric
representation of a general mixed state on the single qubit on site
1 as a point $\vec{a}$ on (or inside) the Bloch sphere \citep{MARINESCU2012221}
\begin{equation}
\rho=\frac{1}{2}\left(\mathbbm{1}+\vec{a}\cdot\vec{\sigma}\right)\,.\label{eq:blochsphere}
\end{equation}
In that representation the trace distance becomes half the Euclidean
distance between the points. The initial state is a pure state on
the unit Bloch sphere, represented by a vector $\vec{a}$ with $\vec{a}^{2}=1$.
As scrambling proceeds, $\rho(t)$ becomes a mixed state represented
by a vector $\vec{b}$ with $\vec{b}^{2}<1$. This implies
\[
D(\rho(t),\rho_{MF}(t))\geq D(\rho_{1}(t),\rho_{MF,1}(t))\geq\frac{1}{2}\left(1-|\vec{b}|\right)
\]
using the triangle inequality to bound the minimum distance between
the points. The purity of \eqref{eq:blochsphere} is
\[
P(t)=\frac{1}{2}\left(1+\vec{b}^{2}\right)
\]
so
\begin{equation}
D(\rho(t),\rho_{MF}(t))\geq\frac{1}{2}\left(1-\sqrt{2P-1}\right)\,.\label{eq:tbound}
\end{equation}
Approximating this for early times when $P$ is close to 1, gives
\[
D(\rho(t),\rho_{MF}(t))\geq\frac{1}{2}\left(1-P\right)\approx\frac{8t^{2}}{3N}\,.
\]

\bibliographystyle{utphys}
\bibliography{BHreference}

\end{document}